\documentclass[prb,showpacs,twocolumn,aps,superscriptaddress,a4paper]{revtex4}
\usepackage{dcolumn}
\usepackage{amsmath}
\usepackage{graphicx}
\usepackage{latexsym}
\usepackage{amsfonts}
\usepackage{amssymb}
\DeclareGraphicsExtensions{.pdf,.gif,.jpg}

\newcommand{\be}{\begin{equation}}
\newcommand{\ee}{\end{equation}}
\newcommand{\beq}{\begin{eqnarray}}
\newcommand{\eeq}{\end{eqnarray}}

\begin{document}
    
\def\bbe{\mbox{\boldmath $e$}}
\def\bbf{\mbox{\boldmath $f$}}    
\def\bg{\mbox{\boldmath $g$}}
\def\bh{\mbox{\boldmath $h$}}
\def\bj{\mbox{\boldmath $j$}}
\def\bq{\mbox{\boldmath $q$}}
\def\bp{\mbox{\boldmath $p$}}
\def\br{\mbox{\boldmath $r$}}    

\def\bone{\mbox{\boldmath $1$}}    

\def\dr{{\rm d}}

\def\tb{\bar{t}}
\def\zb{\bar{z}}

\def\tgb{\bar{\tau}}

\def\bC{\mbox{\boldmath $C$}}
\def\bG{\mbox{\boldmath $G$}}
\def\bH{\mbox{\boldmath $H$}}
\def\bK{\mbox{\boldmath $K$}}
\def\bM{\mbox{\boldmath $M$}}
\def\bN{\mbox{\boldmath $N$}}
\def\bO{\mbox{\boldmath $O$}}
\def\bQ{\mbox{\boldmath $Q$}}
\def\bR{\mbox{\boldmath $R$}}
\def\bS{\mbox{\boldmath $S$}}
\def\bT{\mbox{\boldmath $T$}}
\def\bU{\mbox{\boldmath $U$}}
\def\bV{\mbox{\boldmath $V$}}
\def\bZ{\mbox{\boldmath $Z$}}

\def\bcalS{\mbox{\boldmath $\mathcal{S}$}}
\def\bcalG{\mbox{\boldmath $\mathcal{G}$}}
\def\bcalE{\mbox{\boldmath $\mathcal{E}$}}

\def\bgG{\mbox{\boldmath $\Gamma$}}
\def\bgL{\mbox{\boldmath $\Lambda$}}
\def\bgS{\mbox{\boldmath $\Sigma$}}

\def\bgr{\mbox{\boldmath $\rho$}}

\def\a{\alpha}
\def\b{\beta}
\def\g{\gamma}
\def\G{\Gamma}
\def\d{\delta}
\def\D{\Delta}
\def\e{\epsilon}
\def\ve{\varepsilon}
\def\z{\zeta}
\def\h{\eta}
\def\th{\theta}
\def\k{\kappa}
\def\l{\lambda}
\def\L{\Lambda}
\def\m{\mu}
\def\n{\nu}
\def\x{\xi}
\def\X{\Xi}
\def\p{\pi}
\def\P{\Pi}
\def\r{\rho}
\def\s{\sigma}
\def\S{\Sigma}
\def\t{\tau}
\def\f{\phi}
\def\vf{\varphi}
\def\F{\Phi}
\def\c{\chi}
\def\w{\omega}
\def\W{\Omega}
\def\Q{\Psi}
\def\q{\psi}

\def\ua{\uparrow}
\def\da{\downarrow}
\def\de{\partial}
\def\inf{\infty}
\def\ra{\rightarrow}
\def\bra{\langle}
\def\ket{\rangle}
\def\grad{\mbox{\boldmath $\nabla$}}
\def\Tr{{\rm Tr}}
\def\Re{{\rm Re}}
\def\Im{{\rm Im}}

\title{A time-dependent approach to electron pumping in open quantum systems}

\author{G. Stefanucci}
\email{gianluca@physik.fu-berlin.de}
\affiliation{Institut f\"ur Theoretische Physik, Freie Universit\"at Berlin, 
Arnimallee 14, D-14195 Berlin, Germany}
\affiliation{European Theoretical Spectroscopy Facility (ETSF)}

\author{S. Kurth}
\affiliation{Institut f\"ur Theoretische Physik, Freie Universit\"at Berlin, 
Arnimallee 14, D-14195 Berlin, Germany}
\affiliation{European Theoretical Spectroscopy Facility (ETSF)}

\author{A. Rubio}
\affiliation{European Theoretical Spectroscopy Facility (ETSF)}
\affiliation{Departamento de F\'{i}sica de Materiales, Facultad de Ciencias 
Qu\'{i}micas, UPV/EHU, Unidad de Materiales Centro Mixto CSIC-UPV/EHU 
and Donostia International  Physics Center (DIPC), San Sebasti\'{a}n, Spain}

\author{E. K. U. Gross}
\affiliation{Institut f\"ur Theoretische Physik, Freie Universit\"at Berlin, 
Arnimallee 14, D-14195 Berlin, Germany}
\affiliation{European Theoretical Spectroscopy Facility (ETSF)}

\begin{abstract}
We propose a time-dependent approach to investigate the motion of electrons
in quantum pump device configurations. The occupied one-particle states are propagated in 
real time and used to calculate the local 
electron density and current. An advantage of the present 
computational scheme is that the same computational effort is 
required to simulate monochromatic, polychromatic and nonperiodic 
drivings.  Furthermore, initial state 
dependence and history effects are naturally accounted for. 
This approach can also be embedded 
in the framework of time-dependent density functional theory to 
include electron-electron interactions.
In the special case of periodic drivings we combine the Floquet theory 
with nonequilibrium Green's functions and obtain a general expression 
for the pumped current in terms of inelastic transmission 
probabilities. This latter result is used for benchmarking 
our propagation scheme in the long-time limit. Finally, we discuss 
the limitations of Floquet-based schemes and suggest our approach as 
a possible way to go beyond them. 
\end{abstract}

\date{\today}


\maketitle

\section{Introduction}
\label{intro}

The continuous progress in manipulating single molecules chemically 
bound to macroscopic reservoirs has led to the  
emerging field of molecular electronics.\cite{cfr.2005}
Besides the widely studied stationary case, today 
experimental techniques enable the study of time-dependent phenomena 
in open quantum systems, like photon-assisted transport and electron pumping 
through realistic or artificial molecules.

An electron pump is an electronic device generating a net 
current between two unbiased electrodes. Pumping is achieved by
applying a periodic gate voltage depending on two or more parameters.
Electron pumps have been realized experimentally, e.g., for an open semiconductor 
quantum dot\cite{smcg.1999} driven by two harmonic gate voltages with a phase 
shift, and for a open nanotube\cite{lbtsajwc.2005}  driven by an electrostatic potential 
wave.

In the literature different techniques have been used to discuss electron 
pumping theoretically. For slowly varying electric field the device 
remains in equilibrium and the pumping process is 
adiabatic.\cite{ag.1999,smcg.1999} Brouwer \cite{b.1998} has suggested a 
scattering approach to describe adiabatic pumpings, but his treatment is 
limited to periodic potentials depending on only two parameters.
The generalization to arbitrary periodic potentials has been put 
forward by Zhou et al.\cite{zsa.1999} who used the Keldysh technique to calculate 
the net charge transported across the device per period.

A natural way to go beyond the adiabatic case is to apply Floquet theory. 
Within an equation-of-motion approach Camalet et al.\cite{clkh.2003} 
have found a general expression for the average total 
current and for the noise power of electrons pumped in a tight-binding wire. 
Alternatively, one can combine Floquet theory with non-equilibrium Green's 
function techniques.\cite{sw.1996,a.2005} Generally speaking, Floquet-based approaches 
provide a very powerful tool to calculate average quantities of 
periodically driven systems. However, going beyond the 
monochromatic case quite quickly becomes computationally demanding. 
Furthermore, such approaches  are not 
applicable to the study of transient effects and non-periodic phenomena. 
 
In this work we propose a time-dependent approach suited to study the 
effects of an electric field, like a gate voltage or a laser field, 
on the electron dynamics of a nanoscale junction. Our approach allows
for calculating the full time dependence (including the transient 
behavior) of observables like the local density or current, and
the same computational effort is required for both
monochromatic and polychromatic drivings as well as for nonperiodic 
perturbations. 

The paper is organized as follows. In Section \ref{theory} we 
describe the system consisting of two macroscopic reservoirs connected 
to a central device. We combine the Floquet theory with the Keldysh 
formalism to study the long-time behavior of the device, and we 
generalize the formula for the average current by Camalet et al..\cite{clkh.2003}
Some general features of a Floquet-based algorithm are discussed.
To overcome the 
limitations of the Floquet theory we describe a real-time 
approach based on the propagation of the occupied single-particle 
states. Full implementation details are given for one-dimensional 
electrodes and arbitrary device geometries. The performance of the
algorithm is illustrated in Section \ref{results} where
we specialize to one-dimensional systems and investigate pumping of 
electrons through three different structures: a single barrier, 
a series of barriers and a quantum well. In Section \ref{conc}
we summarize the main results and discuss future projects. 

\section{Time-dependent current}
\label{theory}

We consider an open quantum system $C$ (central region) 
connected to two macroscopically large reservoirs $L$ and $R$ (left 
and right electrodes). We are 
interested in describing the electron dynamics when region $C$ is 
disturbed by arbitrary time-dependent electric fields. Assuming that the 
reservoirs are not directly connected, the one-particle Hamiltonian of the 
entire system reads
\begin{equation}
    \bH(t)=\left[
    \begin{array}{ccc}
	\bH_{LL} & \bH_{LC} & 0 \\
	\bH_{CL} & \bH_{CC}(t) & \bH_{CR} \\
	0 & \bH_{RC} & \bH_{RR}
    \end{array}
    \right].
    \label{ham}
\end{equation}
The Hamiltonian $\bH_{\a\a}$, $\a=L,R$, as well as the Hamiltonian of 
the central region $\bH_{CC}$, are obtained by projecting the full 
Hamiltonian $\bH$ onto the subspace of the corresponding 
region. How to choose the one-particle states in regions $L,\;R$ or $C$ 
depends on the specific problem at hand. We can use, e.g., a real-space grid for 
ab-initio calculations, or a tight-binding representation for model 
calculations, or even different basis functions for different regions 
(for instance, eigenfunctions of the reservoirs for $L$ and $R$, 
and localized states for $C$).\cite{heiko} The off-diagonal parts in Eq. (\ref{ham}) 
account for the contacts and are given in terms of matrix 
elements of $\bH$ between states of $C$ and states of $L$ and $R$. 

In many applications of physical interest the driving field is 
periodic in time. In this case it is possible to work out an analytic 
expression for the dc component of the total current, $I_{\rm dc}$, provided  
memory effects and initial-state dependence are washed out in the 
long time limit. Below we combine the Floquet formalism with nonequilibrium 
Green's functions and generalize the formula for $I_{\rm dc}$ by Camalet et 
al.\cite{clkh.2003} to arbitrary contacts. 
We also discuss the limitations of Floquet theory and propose an 
alternative approach based on the real time propagation of the initially 
occupied states of the system.

\subsection{Long time limit: Floquet theory and Keldysh formalism}
\label{floquet}

Most approaches to driven nanoscale systems are based on a 
fictitious partitioning first introduced by Caroli and 
coworkers.\cite{ccns.1971} The initial many-particle state 
is a Slater determinant of eigenstates of the {\em isolated} left and right 
reservoirs with eigenenergy below some chemical potential $\m$. 
A more physical initial state has been considered by Cini.\cite{c.1980} It is a 
Slater determinant of eigenstates of the {\em contacted} system $L+C+R$ 
with eigenenergy smaller than $\m$. Independently of the initial state, 
it has been proved\cite{sa1.2004,sa2.2004} that the number of 
electrons per unit time that leave the $\a=L,\;R$ reservoir 
is given by the formula\cite{jwm.1994}
\begin{equation}
I_{\alpha}(t)=2\,{\rm Re}\,{\rm Tr}
\left[\bQ_{\a}(t)
\right],
\label{currtt}
\end{equation}
\begin{equation}
    \bQ_{\a}(t)=
    \{\bG^{\rm R}\cdot\bgS^{<}_{\a}+
    \bG^{\rm R}\cdot\bgS^{<}\cdot
    \bG^{\rm A}\cdot\bgS^{\rm A}_{\a}\}(t;t),
    \label{kernel}
\end{equation}
provided that a) $t$ goes to infinity and b) the retarded Green's function 
projected on the central region, $\bG^{\rm R}$, [or the advanced 
one, $\bG^{\rm A}$] vanishes when the separation between its time 
arguments goes to infinity. In the above 
equation $\bgS=\bgS_{L}+\bgS_{R}$ is the embedding self-energy in the 
long time limit 
and the symbol ${\rm Tr}$ denotes a trace over a complete set of 
states of the central region. We also have used the short-hand 
notation $\{f\cdot g\}(t_{1};t_{2})\equiv \int_{-\inf}^{\inf}\dr 
\bar{t}f(t_{1};\bar{t})g(\bar{t};t_{2})$ for the convolution of two 
functions $f$ and $g$.

For an applied bias $U_{\a}$ in reservoir $\a=L,\;R$, which is constant 
in time, the 
embedding self-energy depends only on the difference between its time 
arguments. Let 
\begin{equation}
    \bgS^{\rm R/A}_{\a}(\w)=
    \bgL_{\a}(\w)\mp \frac{i}{2}\bgG_{\a}(\w)
    \label{rase}
\end{equation}
be the Fourier 
transform of the retarded/advanced self-energy. The imaginary part 
$\bgG_{\a}$ is the contribution of region $\a$ to the local spectral density. 
The Fourier transform of the lesser self-energy is then given by 
\begin{equation}
    \bgS^{<}_{\a}(\w)=if_{\a}(\w)\bgG_{\a}(\w),
\end{equation}
where $f_{\a}(\w)=f(\w-U_{\a})$ is the Fermi distribution function.

Let us specialize to periodic time dependent perturbations in 
region $C$: $\bH_{CC}(t)=\bH_{CC}(t+T_{0})$. According to Floquet theory, we assume that the Green's 
function in Eq. (\ref{kernel}) can be expanded as follows
\begin{equation}
\bG^{\rm R/A}(t;t')=\sum_{m}\int\frac{d\w}{2\p}
\bG^{\rm R/A}_{m}(\w)e^{-i\w(t-t')+im\w_{0}t'},
\label{gtira}
\end{equation}
where $\w_{0}=2\p/T_{0}$ is the frequency of the driving field. 
We wish to emphasize that the above expansion is justified only if all  
obervable quantities (calculable from $\bG$) oscillate in time with 
the same frequency as the external field. As pointed out by Hone et 
al.,\cite{hkk.1997} this 
is a questionable assumption.

Inserting Eq. (\ref{gtira}) into Eq. (\ref{kernel}) and extracting the 
dc component we obtain
\begin{eqnarray}
    \bQ_{\a,\rm dc}&\equiv&\lim_{t\ra\inf}\frac{1}{T_{0}}\int_{t}^{t+T_{0}}
    \dr\bar{t}\,\bQ_{\a}(\bar{t})
    \nonumber \\ 
    &=&
    \int\frac{\dr\w}{2\p}
    \bG_{0}(\w)\bgS^{<}_{\a}(\w)+
    \int\frac{\dr\w}{2\p}
    \nonumber \\ 
    &\times&\sum_{m}\bG_{m}(\w)
    \bgS^{<}(\w)\bG^{\dag}_{m}(\w)
    \bgS^{\rm A}_{\a}(\w-m\w_{0}),
    \label{kerneldc}
\end{eqnarray}
where we have defined 
\begin{equation}
    \bG_{m}(\w)\equiv \bG^{\rm R}_{m}(\w-m\w_{0})=
    [\bG^{\rm A}_{-m}(\w)]^{\dag}.
    \label{relation}
\end{equation}
The last equality in Eq. (\ref{relation}) follows directly from the 
identity $\bG^{\rm R}(t;t')=[\bG^{\rm A}(t';t)]^{\dag}$.
The dc component $I_{\a,\rm dc}$ of the time dependent total current 
$I_{\a}(t)$ is given by the right hand side of Eq. (\ref{currtt}) 
with $\bQ_{\a}(t)$ replaced by $\bQ_{\a,\rm dc}$. In Appendix 
\ref{accurr} we show that in the monochromatic case, 
\begin{equation}
    \bH_{CC}(t)=
    \bH_{CC}^{0}+\bU_{+} e^{i\w_{0}t}+\bU_{-}e^{-i\w_{0}t},
\end{equation}
the resulting expression for $I_{\a,\rm dc}$ can be cast in a 
Landauer-like formula 
\begin{equation}
    I_{L,\rm dc}=
    \sum_{m}\int 
    \frac{d\w}{2\p}\left[f_{L}(\w)T_{m,L}(\w)-f_{R}(\w)T_{m,R}(\w)
    \right],
    \label{dcac}
\end{equation}
and $I_{R,\rm dc}=-I_{L,\rm dc}$, as it should be due to charge 
conservation. The ``inelastic'' transmission coefficients $T_{m,\a}$ 
may be interpreted as the probability for electrons to be transmitted from one 
reservoir to the other with the absorption/emission of 
$m$ quanta of the driving field. They can be written as 
\begin{equation}
    T_{m,L}(\w)={\rm 
    Tr}\left[\bgG_{L}(\w)\bG_{m}^{\dag}(\w)\bgG_{R}(\w-m\w_{0})\bG_{m}(\w)\right],
    \label{itpl}
\end{equation}
\begin{equation}
    T_{m,R}(\w)={\rm 
    Tr}\left[\bgG_{R}(\w)\bG_{m}^{\dag}(\w)\bgG_{L}(\w-m\w_{0})\bG_{m}(\w)\right].
        \label{itpr}
\end{equation}

We observe that for zero driving the Fourier coefficients $\bG_{m}$, 
and hence the transmission probabilities $T_{m,\a}$, are all zero 
except for $m=0$, and Eq. (\ref{dcac}) reduces to the well-known Landauer 
formula for steady-state currents.\cite{fl.1981} On the contrary, all the 
$T_{m,\a}$'s contribute to 
the average current when a driving field is present. The 
corresponding $\bG_{m}$'s 
can be computed recursively from the zero-th order coefficient 
$\bG_{0}$, and we defer the reader to Appendix \ref{accurr} for a 
practical implementation scheme. It is also worth emphasizing that 
our formula for the $T_{m,\a}$'s correctly reduces to the one of Camalet 
et al.\cite{clkh.2003}
for a central region described by a tight-binding 
wire of sites $|1\ket,\ldots,|N\ket$ and connected to the left reservoir 
through $|1\ket$ and to the right reservoir through $|N\ket$. In this 
case, the spectral density matrices $\bgG_{\a}$ have 
only one nonvanishing entry, $[\bgG_{L}]_{n,m}=\d_{n,1}\d_{m,1}\g_{L}$ 
and $[\bgG_{R}]_{n,m}=\d_{n,N}\d_{m,N}\g_{R}$, and the coefficients 
$T_{m,\a}$ can be rewritten as
\begin{equation}
    T_{m,L}(\w)=\g_{L}(\w)\g_{R}(\w-m\w_{0})\left|[\bG_{m}(\w)]_{N,1}\right|^{2},
\end{equation}
\begin{equation}
    T_{m,R}(\w)=\g_{R}(\w)\g_{L}(\w-m\w_{0})\left|[\bG_{m}(\w)]_{1,N}\right|^{2}.
\end{equation}

Equation (\ref{dcac}) demonstrates how the initial Floquet assumption 
of Eq. (\ref{gtira}) allows for carrying the analytic calculation of the 
current [Eq. (\ref{currtt})] much further and eventually delivers 
a simple numerical scheme for the computation of the average current. 
Despite the enormous success in predicting ac dynamical properties of 
many different nanoscale conductors,
Floquet theory might be inadequate to face the future challenges of 
nanotechnology.\cite{klh.2005}  Below we discuss some limitations 
of Floquet-based approaches.

(i) {\em Numerical performance}. For later comparison with our 
proposed real time approach, we briefly report on the numerical performance of 
Floquet algorithms, like the recursive scheme proposed in Appendix \ref{accurr}.
Let $N$ be the number of basis functions in region $C$. For a given 
frequency $\w$ the calculation of $\bG_{0}(\w)$ requires the inversion 
of $m_{\rm max}$ complex matrices of dimension $N\times N$. The 
number $m_{\rm max}$ should be chosen such that the 
cut-off energy $E_{\rm max}=m_{\rm max}\w_{0}$ is much larger than 
any other energy scale in the problem. Typically, $m_{\rm max}$ is 
in the range from $10$ to $100$. The coefficients $\bG_{\pm m}(\w)$, $m>0$, are then 
calculated from $\bG_{\pm (m-1)}(\w)$ by simple matrix multiplications 
according to Eq. (\ref{recgm}). Knowing the $\bG_{m}$'s one can compute 
the inelastic transmission probabilities from Eqs. 
(\ref{itpl},\ref{itpr}), and hence the average current. 

In the above procedure most of the computational time is spent 
for matrix inversions and matrix multiplications. We can roughly extimate the overall time of 
a full run as $T_{\rm run}\simeq  m_{\rm max}\times N_{\w}\times 
(\t_{i}+\t_{m})$, where $N_{\w}$ is the number of mesh points 
(generally not uniform) along the 
$\w$ axis used to evaluate the integral in Eq. (\ref{dcac}), and
$\t_{i}$ ($\t_{m}$) is the time for a 
single matrix inversion (multiplication). In our case both
$\t_{i}$ and $\t_{m}$ scale as $N^{3}$.
Depending on the system and on the external driving forces, the 
inelastic transmission probabilities might exhibit quite sharp peaks 
as function of energy. Therefore for an accurate computation of the energy 
integral in Eq. (\ref{dcac}) a fine energy grid is required, 
which means that $N_{\w}$ is large. In the numerical 
calculations of Section \ref{results} $N_{\w}$ is in the range 
$100$ to $1000$. We conclude that $T_{\rm run}/(\t_{i}+\t_{m}) \sim 
10^{3}$ to $10^{5}$. 

(ii) {\em Periodic potentials}.  Beyond the monochromatic case,
the recursive scheme of Appendix \ref{accurr} becomes computationally 
demanding. 
The inclusion of one, two, \ldots more harmonics in the expansion of 
the driving field [see Eq. (\ref{uexp})] transforms the 
block tridiagonal system of equations for the $\bG_{m}$'s into a 
block penta-diagonal, hepta-diagonal, \ldots system of equations. 
For arbitrary periodic drivings a Floquet-based approach
may not be feasible.

(iii) {\em Arbitrary time dependent potentials}.
Besides the wide class of periodic drivings, it is of interest  
to investigate the response of a nanodevice to non-periodic 
drivings as well.\cite{oaf.2005} In such cases the Floquet formalism does not apply 
and a full time-dependent approach is required.

(iv) {\em Transients}. The Landauer formalism provides a very powerful 
technique to calculate non-equilibrium quantities in steady-state 
regimes. Similarly, the Floquet formalism allows to calculate 
non-equilibrium quantities in ``oscillating-state'' regimes, i.e., 
when all transient effects are died off. However, transient responses 
can be expected to become of some relevance in the future. Molecular 
devices will eventually be integrated in nanoscale circuits and 
respond to ultrafast external signals. Transient effects in such 
operative regimes may not be irrelevant, as it has been recently 
recognized by several 
authors.\cite{sa1.2004,svk.2005,ksarg.2005,vsa.2006,sssbht.2006,mwg.2006,bsdv.2005}
In Section \ref{results} we provide explicit evidence of long-lived 
superimposed oscillations in the time-dependent current profile.
The frequencies of these oscillations are not commensurable with 
the driving frequency, and have to be ascribed to the presence of 
``adiabatic'' bound states.\cite{ds.2006,s.2006}

\subsection{Real time propagation}
\label{prop}

In this Section we propose an alternative approch to driven nanoscale transport. 
The main idea is to calculate the time-dependent total current from 
the time-dependent wavefunctions $|\q_{s}(t)\ket$, where $|\q_{s}(0)\ket$ 
is the $s$-th eigenstate of the system $L+C+R$ before the 
time-dependent perturbation is switched on. Our approach 
does not rely on the Floquet assumption, and is 
free from {\em all} the limitations discussed previously. Furthermore, the 
computational time is comparable with Floquet-based algorithms.

As the full Hamiltonian $\bH(t)$ refers to an extended and non-periodic 
system, we cannot solve {\em brute force} the Schr\"odinger equation 
\begin{equation}
    i\frac{\dr}{\dr t}|\q(t)\ket = \bH(t) |\q(t)\ket.
    \label{tdse}
\end{equation}
Fortunately, we do not need to calculate the 
time dependent wavefunction everywhere in the system in order 
to calculate the total current. The knowledge of the wavefunction in 
region $C$ is enough for our purposes (see below). Denoting with $|\q_{C}(t)\ket$ 
the wavefunction projected on region $C$ and with $|\q_{\a}(t)\ket$ 
the wavefunction projected on region $\a=L,\; R$, it is 
straightforward to show that Eq. (\ref{tdse}) implies the following 
equation for $|\q_{C}(t)\ket$\cite{hf.1994}
\begin{eqnarray}
    i\frac{\dr}{\dr t}|\q_{C}(t)\ket&=&\bH_{CC}(t)|\q_{C}(t)\ket
    +\int_{0}^{t}\bgS^{\rm R}(t;t')|\q_{C}(t')\ket
    \nonumber \\ 
    &+&\sum_{\a=L,R}\bH_{C\a}\exp(-i\bH_{\a\a}t)|\q_{\a}(0)\ket,
    \label{tdsec}
\end{eqnarray}
where 
\begin{equation}
    \bgS^{\rm R}(t;t')=
    \sum_{\a=L,R}\bH_{C\a}\exp\left(-i\bH_{\a\a}(t-t')\right)\bH_{\a C}
\end{equation}
is the Fourier transform of the embedding self-energy in 
Eq. (\ref{rase}). 

Equation (\ref{tdsec}) is an {\em exact} equation for the time 
evolution of open systems, but is still not 
suited for a numerical implementation. The importance of charge 
conservation in quantum transport makes the unitary property 
a fundamental requirement. In this work we use a unitary algorithm which 
has been recently 
proposed to study electron transport in biased electrode-device-electrode 
systems.\cite{ksarg.2005} Below we illustrate the main ideas and 
specialize the formulas of Ref. \onlinecite{ksarg.2005} to 
one-dimensional reservoirs. 

For a given initial state $|\q(0)\ket=|\q^{(0)}\ket$ we calculate the 
time-evolved state $|\q(t_{m}=2 m \d)\ket=|\q^{(m)}\ket$ by approximating 
Eq. (\ref{tdse}) with the Crank-Nicholson formula
\begin{equation}
    \left({\bf 1}+i\d\bH^{(m)}\right)|\q^{(m+1)}\ket
    =\left({\bf 1}-i\d\bH^{(m)}\right)|\q^{(m)}\ket,
    \label{cn}
\end{equation}
with $\bH^{(m)}=\frac{1}{2}[\bH(t_{m+1})+\bH(t_{m})]$. The above 
propagation scheme is unitary (norm conserving) and accurate to 
second-order in $\d$. From Eq. (\ref{cn}) we can extract an equation 
for the time-evolved state in region $C$, similarly to what we have 
done for the derivation of Eq. (\ref{tdsec}). The final result is
\begin{eqnarray}
    \left({\bf 1}_{C}+i\d\bH_{\rm eff}^{(m)}\right)|\q_{C}^{(m+1)}\ket&=&
    \left({\bf 1}_{C}-i\d\bH_{\rm eff}^{(m)}\right)|\q_{C}^{(m)}\ket
    \nonumber \\ &+&
    |S^{(m)}\ket-|M^{(m)}\ket,
    \label{cnc}
\end{eqnarray}
with ${\bf 1}_{C}$ the identity matrix in region $C$. Equation 
(\ref{cnc}) is the proper (unitary) time-discretization of Eq. 
(\ref{tdsec}). Moreover,  Eq. (\ref{cnc}) is ready to be implemented 
since it contains only {\em finite-size} matrices and vectors (with the  
dimension used to describe the central region as, e.g., the number of 
lattice sites in a tight-binding representation or the number of grid 
points in a real-space grid representation). 
In the following we give full implementation 
details of the various terms in Eq. (\ref{cnc}).

\begin{figure}[htbp]
\includegraphics*[width=.46\textwidth]{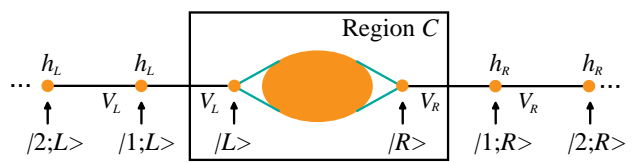}
\caption{The schematic sketch of the electrode-junction-electrode 
system with semiperiodic one-dimensional electrodes. }
\label{system}
\end{figure}

For the sake of simplicity, we consider one-dimensional semi-infinite 
reservoirs described by tridiagonal matrices $\bH_{\a\a}$, 
$\a=L,\; R$, with diagonal entries $h_{\a}$ and off-diagonal entries 
$V_{\a}$, see Fig. \ref{system}. For tight-binding models, the parameter $h_{\a}$ represents 
the on-site energy while the parameter $V_{\a}$ represents the hopping 
integral between 
nearest-neighbour sites. The Hamiltonian $\bH_{\a\a}$ is also 
suited to describe continuum models with a three-point discretization of 
the kinetic term. In this case, the parameter $h_{\a}=1/\D 
x^{2}+U_{\a}$ and $V_{\a}=-1/(2\D x^{2})$, where $\D x$ is the grid spacing. 
We would like to emphasize that the algorithm can easily be generalized to 
reservoirs with an arbitrary semi-infinite periodicity and it is not 
limited to one-dimensional systems.\cite{ksarg.2005}

Without loss of generality, we consider a central region that 
includes few sites of the left and right reservoirs, and we denote by 
$|\a\ket$ the state where only the site 
of region $C$ connected to the reservoir $\a=L,\; R$ is occupied (see 
Fig. \ref{system}).

The {\em memory state} $|M^{(m)}\ket$ 
stems from the the second term on the r.h.s. of Eq. (\ref{tdsec}) and 
reads
\begin{equation}
    |M^{(0)}\ket =0,
\end{equation}
while for $m\geq 1$ we have
\begin{eqnarray}
    |M^{(m)}\ket&=&\d^{2}\sum_{\a=L,R}|\a\ket\sum_{k=0}^{m-1}
    \left[
    \bra\a|\q_{C}^{(k+1)}\ket+\bra\a|\q_{C}^{(k)}\ket
    \right]
    \nonumber \\
    &\times&
    \left[q_{\a}^{(m-k)}+q_{\a}^{(m-k-1)}\right].
\end{eqnarray}
The $q$-coefficients can be computed recursively according to
\begin{equation}
    q_{\a}^{(0)}=\frac{-(1+i\d h_{\a})+\sqrt{(1+i\d h_{\a})^{2}+
    (2\d V_{\a})^{2}}}{2\d^{2}},
    \label{q0}
\end{equation}
\begin{equation}
    q_{\a}^{(1)}=\frac{1-i\d h_{\a}-2\d^{2}q_{\a}^{(0)}}{1+i\d 
    h_{\a}+2\d^{2}q_{\a}^{(0)}}q_{\a}^{(0)},
    \label{q1}
\end{equation}
and for $m\geq 2$
\begin{eqnarray}
q_{\a}^{(m)}&=&\frac{q_{\a}^{(1)}q_{\a}^{(m-1)}}{q_{\a}^{(0)}}-
\d^{2}\frac{q_{\a}^{(0)}q_{\a}^{(m-2)}}{1+i\d h_{\a}+2\d^{2}q_{\a}^{(0)}}
\nonumber \\ 
&-&\d^{2}\sum_{k=1}^{m-1}\frac{q_{\a}^{(k)}+2q_{\a}^{(k-1)}+q_{\a}^{(k-2)}
}{1+i\d h_{\a}+2\d^{2}q_{\a}^{(0)}}q_{\a}^{(m-k)},
\label{qqm} 
\end{eqnarray}
with the convention that $q_{\a}^{(m)}=0$ for negative $m$. 

The {\em source state} $|S^{(m)}\ket$ 
stems from the last term on the r.h.s. of Eq. (\ref{tdsec}) and 
reads
\begin{equation}
    |S^{(m)}\ket=-2 i \d \sum_{\a=L,R}
    \bH_{C\a}\frac{({\bf 1}_{\a}-i\d\bH_{\a\a})^{m}}
    {({\bf 1}_{\a}+i\d\bH_{\a\a})^{m+1}}|\q_{\a}(0)\ket,
\end{equation}
where ${\bf 1}_{\a}$ is the unit matrix in region $\a$. The source 
state depends on the initial wavefunction in the 
reservoirs. As we are interested in propagating 
eigenstates of $\bH(0)$, $|\q_{\a}(0)\ket$ 
has the following general expression
\begin{equation}
    |\q_{\a}^{(0)}\ket=
    A^{(+)}_{\a}|+p_{\a}\ket+A^{(-)}_{\a}|-p_{\a}\ket,
    \label{asympsi}
\end{equation}
with
\begin{equation}
    | p_{\a}\ket=\sum_{j=1}^{\inf}e^{ i p_{\a} j}|j;\a\ket,
\end{equation}
and the state $|j;\a\ket$ where only the $j$-th site of reservoir 
$\a=L,\; R$ is occupied, see Fig. \ref{system}. 
For extendend states in region 
$\a$ the parameter $p_{\a}$ is real. 
For bound states or fully reflected states in region $\a$ 
the parameter $p_{\a}$ is imaginary and the amplitude ($A^{(+)}_{\a}$ or 
$A^{(-)}_{\a}$) of the growing exponential is zero. 
No matter if $p_{\a}$ is real or imaginary, one 
can prove that
\begin{equation}
    \bH_{C\a}\frac{({\bf 1}_{\a}-i\d\bH_{\a\a})^{m}}
    {({\bf 1}_{\a}+i\d\bH_{\a\a})^{m+1}}| p_{\a}\ket = \z_{\a}^{(m)} |\a\ket,
\end{equation}
with
\begin{equation}
    \z_{\a}^{(m)}=e^{ip_{\a}}
    V_{\a}\g_{\a}^{(m)}+i\d 
    \sum_{k=0}^{m} 
    \g_{\a}^{(m-k)}\left[q_{\a}^{(k)}+q_{\a}^{(k-1)}\right],
\end{equation}
and
\begin{equation}
\g_{\a}^{(m)}=\frac{(1-i\d h_{\a}-2i\d V_{\a}\cos p_{\a})^{m}}
{(1+i\d h_{\a}+2i\d V_{\a}\cos p_{\a})^{m+1}}.
\end{equation}

Finally, the {\em effective Hamiltonian} is given by 
\begin{equation}
    \bH_{\rm eff}^{(m)}=
    \bH_{CC}^{(m)}-i\d\sum_{\a=L,R}q_{\a}^{(0)}|\a\ket\bra\a|.
    \label{effham}
\end{equation}

The above algorithm allows us to calculate the time evolution of {\em 
any} initial state whose wavefunction in the reservoirs has the 
form in Eq. (\ref{asympsi}). This is the case of both 
the contacting approach by Caroli and coworkers and the partition-free 
approach by Cini. In the former approach, the initial one-particle 
states are eigenstates of the isolated left and right reservoirs, 
meaning that
\begin{equation}
    |\q_{\a}^{(0)}\ket=2\sum_{j=1}^{\inf}\sin (p_{\a}j)|j;\a\ket=
    \frac{|+p_{\a}\ket-|-p_{\a}\ket}{i},
\end{equation}
for $\a=L$ (or $\a=R$), zero for $\a=R$ (or $\a=L$), and zero in 
region $C$. In the latter approach the computation of the initial 
one-particle states is more involved. Here we have used a recently proposed 
general scheme based on the diagonalization of the imaginary part of the 
retarded Green's function.\cite{ksarg.2005} This scheme may also be used for 
arbitrary, semiperiodic electrodes. In the special case 
of spatially uniform one-dimensional reservoirs one can, of course, always 
use the textbook procedure of matching the wavefunction at the interfaces. 

Denoting with $|\q_{s,C}(t)\ket$ the evolution of the original eigenstate 
$|\q_{s}(0)\ket$ in the central region, 
we can calculate the time dependent occupation $\r(j,t)$
of a state $|j\ket$ in region $C$ according to
\begin{equation}
    \r(j,t)=\sum_{s}f(\ve_{s})\left|\bra j|\q_{s,C}(t)\ket\right|^{2},
    \label{tddens}
\end{equation}
where $\ve_{s}$ is the eigenvalue of $|\q_{s}(0)\ket$ and 
$f(\ve)$ is the Fermi distribution function. 
Similarly, the total time-dependent current $I_{\a}(t)$ can be 
calculated from the time-derivative of the total number of particles 
in electrode $\a$ and reads
\begin{eqnarray}
    I_{\a}(t)&=&-2\sum_{s}f(\ve_{s})\sum_{j\neq \a}
    \nonumber \\ &\times& {\rm Im}\,
    \bra j|\q_{s,C}(t)\ket\bra\q_{s,C}(t)|\a\ket
    [\bH_{CC}(t)]_{\a j},
    \label{tdcurr}
\end{eqnarray}
where the sum is over all states $j$ of region $C$ except the 
state $|\a\ket$.

We wish to conclude this Section with a discussion on the performance of our method 
and a comparison with Floquet-based approaches.

(i) The computational time $T_{\rm run}$ scales linearly with the 
number of states $N_{s}$ used to evaluate the sum in Eq. 
(\ref{tddens}) or Eq. (\ref{tdcurr}), and quadratically with the 
number of time steps $N_{t}$. In most cases transient effects 
disappear after few femtoseconds (few tens of atomic units). 
Using a time step of the order of $10^{-2}$ a.u. we can obtain a rather 
good estimate of $I_{\a,\rm dc}$ with $N_{t}\sim 10^{3}$ to $10^{4}$. 
Given a central regions with  
hundreds of states the real time algorithm can be of the same 
speed of or even faster than 
the Floquet algorithm of Appendix \ref{accurr}.

(ii) The real-time algorithm can deal with arbitrary (periodic and 
non-periodic) drivings, and the computational time is independent of 
the specific time dependence of $\bH_{CC}(t)$. Moreover, the algorithm 
is easily generalized\cite{ksarg.2005} to deal with spatially uniform 
bias potentials in the electrodes with arbitrary dependence on time 
such as, e.g., for an ac bias.

(iii) From the time-evolved states $|\q_{s}(t)\ket$ we have access to 
the total current $I_{\a}(t)$ at {\em any} time $t$, and not only to 
the long-time limit of the dc component of $I_{\a}(t)$. In particular, we 
can easily investigate transients and the full shape of $I_{\a}(t)$ 
for $t\ra\inf$. In practice, this limit is achieved for a finite time 
$T_{\rm max}$ after which all transient phenomena have died out.

(iv) Another advantage of our method is the possibility of including 
electron-electron interactions via time-dependent density functional 
theory.\cite{rg.1984} Indeed, the external potential is local in both space and time 
provided the initial state is the ground state of the contacted 
system. Therefore, according to the Runge-Gross 
theorem,\cite{rg.1984,vl.1999} the interacting time-dependent density can be 
reproduced in a fictitious system of non-interacting electrons moving 
under the influence of an effective Kohn-Sham potential which is 
local in space and time.
We observe that this is not the case in the contacting 
approach since the switching of the contacts makes the external 
potential non-local in space and hence the Runge-Gross theorem does 
not apply.

(v) Finally, we would like to stress that the Hamiltonian $\bH_{CC}(t)$ 
enters in the algorithm only via the effective Hamiltonian $\bH_{\rm 
eff}$ of Eq. (\ref{effham}), and has no restrictions. Thus, besides 
one-dimensional structures (like those considered in Section 
\ref{results}) one can 
consider other geometries as well, like those of planar molecules, 
quantum rings, nanotubes, jellium slabs, etc.

\section{Numerical results}
\label{results}

In this Section we illustrate the performance of the proposed scheme 
by presenting our results for
one-dimensional continuos systems described by the time-dependent Hamiltonian 
\begin{equation}
    H(x,t)=-\frac{\nabla^{2}}{2}+U(x,t).
\end{equation}
We discretize $H$ on a equidistant grid and 
use a three-point discretization for the kinetic term. Within this 
model we study various model systems highlighting different features 
in electron pumping.

\subsection{Archimedean screw}
\label{archscrew}

\begin{figure}[htbp]
\includegraphics*[width=.46\textwidth]{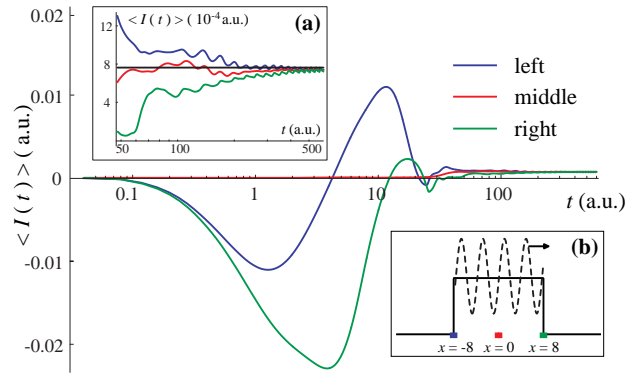}
\caption{Time-dependent average current at the left and 
right interface and in the middle of region $C$ for pumping 
through a single square barrier by a travelling wave. 
The travelling potential wave is restricted to the propagation 
window $|x|< 8$ a.u. and has the form $U(x,t) = U_1 \sin(q x - 
\omega_{0} t)$ with 
amplitude $U_1=0.35$ a.u., wave number $q=1.6$ a.u. and frequency 
$\omega_{0}=0.2$ a.u.. Inset (a) is a magnification of the long-time 
behavior. The straight line corresponds to the value
$I_{L,\rm dc}=7.63\cdot 10^{-4}$ a.u. of the 
average current calculated using the Floquet algorithm. Inset (b)
displays the static potential barrier 
(solid line) and the superimposed right-moving travelling wave (dashed 
line).}
\label{currave}
\end{figure}

As a first example of electron pumping we have calculated the time evolution 
of the density and total current for a single square barrier exposed 
to a travelling potential wave $U(x,t) = U_1 \sin(q x - \omega_{0} t)$. 
The wave is spatially restricted to the explicitly treated device region 
which in our case also coincides with the static potential barrier. 
The barrier extends from $x=-8$ to $x=+8$ a.u. and its height is 0.5 
a.u., see inset (b) in Fig \ref{currave}. The system is unbiased, i.e., 
$U_{L}=U_{R}=0$, and the Fermi energy of the initial (ground) state is 
$\ve_{\rm F}=0.3$ a.u.. For the numerical implementation we have 
chosen a lattice spacing $\D x=0.08$ a.u., and 200 $k$-points between 0 
and $k_{\rm F}=\sqrt{2\ve_{\rm F}}$ which amounts to the propagation 
of 400 states. 

In Fig. \ref{currave} we plot the time-dependent average current calculated 
according to
\begin{eqnarray}
    \bra I(t)\ket&=&\theta(T_{0}-t)\frac{1}{t}\int_{0}^{t}\dr \t \, I(\t)
    \nonumber \\ 
    &+&\theta(t-T_{0})\frac{1}{T_{0}}\int_{t-T_{0}}^{t}\dr \t \, I(\t),
    \label{tdave}
\end{eqnarray}
with the period of the travelling wave $T_{0}=2\p/\w_{0}$. For the 
time propagation we have chosen a time step $2\d=0.02$ a.u.. 
As expected $\bra I(t)\ket$ converges to some steady value 
$I_{L,\rm dc}$ after a transient time of the order of $50\div 60$ a.u.. 
We have calculated the average 
current in three different points of region $C$ and verified that the 
steady value does not depend on the position. The 
dc limit $I_{L,\rm dc}$ can also be computed from the
Floquet algorithm of Appendix \ref{accurr}. Using $m_{\rm max}=15$ 
and $N_{\w}=150$ energy points between 0 and $\ve_{\rm F}$, we find 
$I_{L,\rm dc}=7.63\cdot 10^{-4}$ a.u., in very good agreement with the average current of 
the time propagated system, see inset (a) of Fig. \ref{currave}.

\begin{figure}[htbp]
\includegraphics*[width=.48\textwidth]{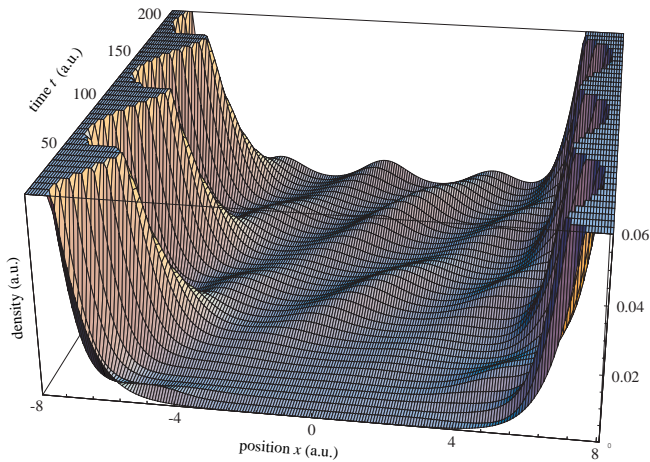}
\caption{Time-dependent density in region $C$ as a function of 
position $x$ and time $t$. The range for $\r(x,t)$ is between 0 and 
0.06 a.u. for improved visibility of the pockets of density. All parameters are the 
same as for Fig. \ref{currave}.}
\label{screw}
\end{figure}

In Fig. \ref{screw} we plot the time-dependent density $\r(x,t)$ in 
the device region as a function of both position $x$ and time $t$.
The density exhibits local maxima in the potential minima 
and is transported in pockets by the wave. From Fig. \ref{screw} it 
is also evident that the height of the pockets is not uniform over the 
system, and reaches its maximum around $x=0$. 
We also notice that the particle 
current flows in the same direction as the driving wave.
The pumping mechanism in this example
resembles pumping of water with the Archimedean screw. 

\subsection{Pumping through a semiconductor nanostructure}
\label{nanopump}

\begin{figure}[htbp]
\includegraphics*[width=.47\textwidth]{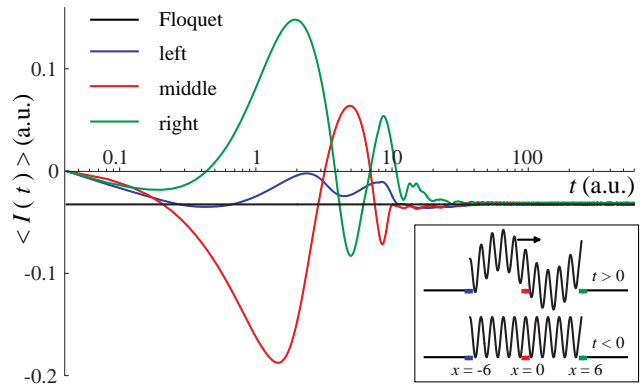}
\caption{Time-dependent average current at the left and 
right interface and in the middle of region $C$ for pumping 
through a device region which extends from $x=-6$ to $x=6$ a.u.. A 
travelling wave $U(x,t) = U_1 \sin(q x - \omega_{0} t)$ with $U_{1}=0.5$ a.u., 
$\w_{0}=0.8$ a.u., and $q=0.6$ a.u., is superimposed to 
the static potential $U_{0}(x)=U_{C}(1+\cos(kx))$  
with $U_{C}=0.5$ a.u. and $k=10\p/6\sim 5.2$ a.u., and the all system 
is unbiased, see inset. 
The straight line corresponds to the value of the 
average current as obtained from the Floquet algorithm which
yields $I_{L,\rm dc}=3.26\cdot 10^{-2}$ a.u..}
\label{waveneg}
\end{figure}

The second example was motivated by a recent 
e\-x\-p\-e\-r\-i\-m\-e\-n\-t on pumping through a carbon nanotube.\cite{lbtsajwc.2005}
The arrangement has been suggested by Talyanskii et al.\cite{tnsl.2001} 
and is as follows. A semiconducting nanotube lying on a quartz substrate is placed
between two metallic contacts. A transducer 
generates an acoustic wave on the surface of the piezoelectric 
crystal. The crystal responds by generating an electrostatic potential 
wave which acts like our travelling wave on the electrons in the 
nanotube. The direction of the pumping current is found to 
depend on the applied gate voltage. A pumping current flowing in 
the direction opposite to the propagation direction of the travelling wave 
has been interpreted in a stationary picture as 
a predominant hole tunneling over electron tunneling.
To reproduce such an inversion in the current flow we
have modelled the nanotube with a periodic static corrugation 
$U_{0}(x)=U_{C}(1+\cos(kx))$ in region $C$, with $U_{C}=0.5$ a.u. and 
$k=10\p/6\sim 5.2$ a.u. 
(see inset in Fig \ref{waveneg}).
For a travelling wave $U(x,t)=U_{1}\sin(qx-\w_{0} t)$, with $U_{1}=0.5$ a.u., 
$\w_{0}=0.8$ a.u., and $q=0.6$ a.u., we have found that the minimum current occurs 
at $\ve_{\rm F}=3.0$ a.u.. All parameters in this example have been 
chosen to better illustrate and discuss the effect of the 
current inversion. The present Section is not intended to give a 
realistic description of some specific experiment. 

In Fig. \ref{waveneg} we plot the time-dependent average current [see Eq. 
(\ref{tdave})] in three different points of the device region. 
For the numerical propagation we have used a lattice spacing $\D x=0.06$ a.u., 
a time step $2\d=0.02$ a.u., and $400$ $k$-points between $0$ and $k_{\rm 
F}=\sqrt{2\ve_{\rm F}}$. The 
system responds to a right-moving travelling wave by generating a net 
current flowing to the left. Again we observe that the transient time 
is of the order of few tens of atomic units, and that the steady value 
is independent of the position. As in the previous example, we used 
the Floquet algorithm of Appendix \ref{accurr} for benchmarking our 
real-time propagation algorithm. Due 
to the high Fermi energy the calculation was carried out with $m_{\rm 
max}=15$ and $N_{\w}=400$ energy points between 0 and $\ve_{\rm F}$.
The result $I_{L,\rm dc}=3.26\cdot 10^{-2}$ a.u. 
is displayed in Fig. \ref{waveneg} with a straight 
line and is in extremely good agreement with the long-time limit 
of the average current obtained from direct propagation in time.

To understand how the electron fluid moves when the direction of the 
current is opposite to that of the 
driving potential wave, we have studied the dynamical flow pattern of the 
density. We emphasize that such a study would have been rather 
complicated in Floquet-based approaches. The latter are often used as ``black boxes'', 
and one needs to resort to limiting cases, like, e.g., 
the adiabatic picture, the high frequency limit, the theory of linear response, 
etc., for a qualitative understanding.

\begin{figure}[htbp]
\includegraphics*[width=.47\textwidth]{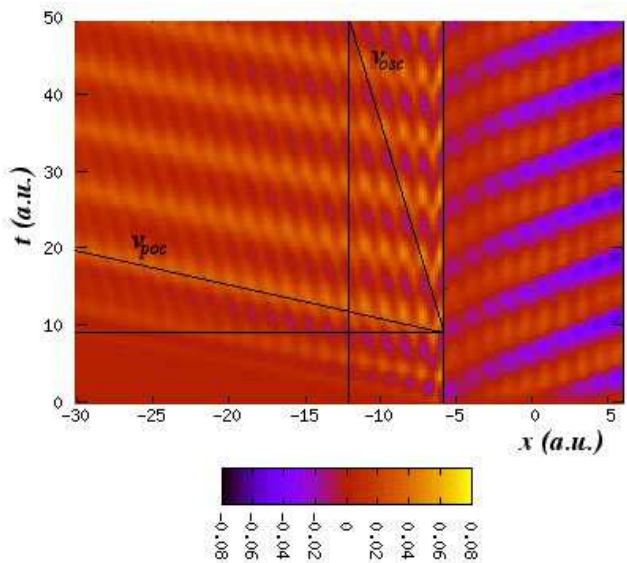}
\caption{Contour plot of the excess density $\D \r(x,t)$ in the device region ($x$ 
between -6 and 6 a.u.) and in a portion of the left reservoir ($x$ between 
-30 and -6 a.u.). Due to the large oscillations of the excess density in the 
device region, $\D \r(x,t)$ has been scaled down by a factor of 10 
for $|x|<6$ a.u.. We also draw straight lines to show a pocket 
trajectory and the trajectory of a superimposed oscillation. All 
parameters are the same as for Fig. \ref{waveneg}. }
\label{excess}
\end{figure}

In Fig. \ref{excess} we display a contour plot of the excess density 
$\D \r(x,t)=\r(x,t)-\r(x,0)$ in an extended region which includes the device region 
and a portion of the left reservoir. In the device region we clearly 
see pockets that are dragged by the travelling 
wave and are moving to the right. However, every pocket with a slightly positive 
$\D \r$ is 
followed by a pocket with a noticeably negative $\D \r$, and the net 
excess density is negative. On the other hand, in the left reservoir 
only pockets with a positive $\D \r$ exist and they move to the left. 
We conclude that the driving produces right-moving ``bubbles'' in the device 
region and that to each bubble corresponds a more dense region of 
fluid moving to the left. One can estimate the speed $v_{\rm poc}$ of the travelling 
pockets from the slope of the patterns at constant density and find 
$v_{\rm poc}\sim \w_{0}/q\sim 1.33$ a.u., as expected. We also notice superimposed 
density oscillations on each pocket. These oscillations have the 
same spatial period of the static corrugation in the device region, and 
move in the same direction of the pockets at a constant speed 
$v_{\rm osc}\sim \w_{0}/k\sim 0.15$ a.u..

\begin{figure}[htbp]
\rotatebox{270}{\includegraphics[width=.355\textwidth]{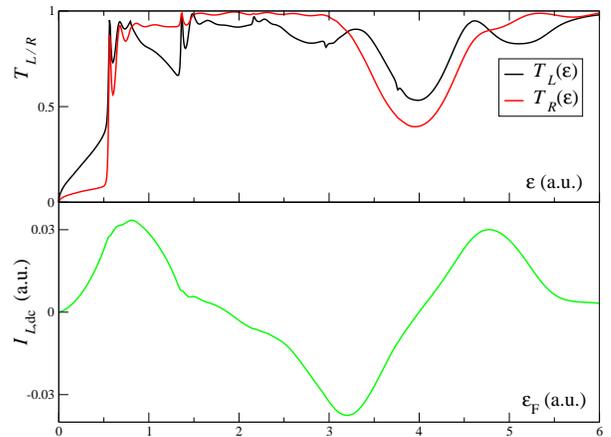}}    
\caption{dc component of pump current $I_{L,\rm dc}$ and left/right transmission 
probabilities $T_{L/R}$ as function of the Fermi energy. The curves have been 
obtained using the Flocquet algorithm of Appendix \ref{accurr} with 
$m_{\rm max}=15$ and 
$N_{\w}=800$ energy points between 0 and $\ve_{\rm F}=6$ a.u..}
\label{somelight}
\end{figure}

In Fig. \ref{somelight} we illustrate  how the pumped current in this model 
depends on the Fermi level. For Fermi energies comparable to the amplitude 
of the corrugated potential in the device region the pumping current is always 
positive, i.e., follows the propagation of the perturbed wave. 
However, there are striking effects that are more or less independent of the 
strength of the perurbation: the pumping current reaches a maximum positive 
value at $\ve_{\rm F}\sim \w_{0} = 0.8$ a.u., then decreases with 
increasing Fermi energy (with the turning point to negative values just  below 
$\ve_{\rm F}=2$ a.u.) and reaches a minimum (negative) value above 
$\ve_{\rm F}=3$ a.u.. To rationalise this behavior we have calculated 
the total transmission probabilities $T_{\a}=\sum_{m}T_{m,\a}$, 
$\a=L,R$, for left- and right-going electrons [see Eqs.~(\ref{itpl}, 
\ref{itpr})]. As 
one can see from Fig. \ref{somelight}, both $T_{L}$ and $T_{R}$ remain quite 
small for Fermi energies below $\D\sim 0.54$ a.u., which roughly corresponds 
to the bottom of the lowest band of the periodic structure of the device. 
In this energy window transport is dominated by tunneling and the pumping 
current follows the travelling wave ($T_{L}>T_{R}$) similar to the case of the 
Archimedean screw, see Section \ref{archscrew}. For $\ve_{\rm F}>\D$ we 
enter the region of resonant transport (the energy of the lowest band) and 
$T_{L}$, $T_{R}$ sharply increase. We observe that for $\ve_{\rm F}< 
\w_0=0.8$ a.u. both 
$T_L$ and $T_R$ have a structure similar to the total transmission function of 
the static case. For $ \w_0 < \ve_{\rm F} < \w_0 + \D$, however, $T_L$ 
decreases significantly while $T_R$ remains fairly constant around 1. We 
interpret this in the following way. The probability of the right-going 
electrons of emitting a photon of frequency $\w_0$ (and therefore reducing 
their energy) is larger than for the left-going electrons. Loosing this 
energy, the transmission $T_L$ resembles the static transmission function 
for energy $\ve_{\rm F} -\w_0$ which has a much lower value. The asymmetry 
between left- and right-going states can easily be understood by realizing 
that the pump wave introduces a preferential direction in the problem. 
As further evidence to support this interpretation we note that for 
$\ve_{\rm F} = \w_0 + \D$ the transmission function $T_L$ increases rapidly 
as for $\ve_{\rm F}= \D$. This can be viewed as a replica of the static 
transmission function shifted by one quantum of energy $\w_0$. Throughout 
the energy window of the lowest band, $T_L$ remains lower than $T_R$. 
As a consequence the pumping current decreases monotonically. This behaviour 
is reversed when the Fermi energy hits the top of the lowest band, around 
$3.4$ a.u.. In the gap (of about $2U_C=1$ a.u.) both $T_{L}$ and $T_{R}$ 
drop and transport is dominated by tunneling again. In this region 
$T_{L}>T_{R}$ and the pumping current increases.

The present model gives positive and negative pumping current as a function of 
the Fermi energy and provides a simple physical interpretation of the 
effect of current inversion. Our picture, however, is somewhat 
different from the one given by Leek et al.. \cite{lbtsajwc.2005} Indeed, in 
their explanation the sign of the pumping current is independent of the 
frequency $\w_0$ of the travelling wave. 
On the other hand, in our case, if the frequency exceeds the width of the 
lowest band, the right-going electrons cannot emit a photon and current 
inversion is not guaranteed anymore.

\subsection{Transients effects}

As a last example we study electron pumping in quantum wells. We will show 
the presence of long-lived superimposed oscillations whose frequency 
is generally not commensurable with the driving frequency. 
The quantum well is modelled with a static 
potential $U_{0}(x)=-1.4$ a.u. for $|x|<1.2$ a.u. 
and zero otherwise. Initially
the system is in the ground state with Fermi 
energy $\ve_{\rm F}=0.1$ a.u.. The unperturbed Hamiltonian has 
two bound-state eigensolutions with energy $\ve^{0}_{b,1}=-1.035$ a.u. and 
$\ve^{0}_{b,2}=-0.156$ a.u.. The ground-state Slater determinant
contains all extended states with energy between 0 and 
$\ve_{\rm F}$ and two localized states with negative energy.
At positive times a constant bias 
$U_{R}=0.1$ a.u. is applied on the right lead and a travelling wave 
$U(x,t)=U_{1}\sin(qx-\w_{0} t)$, with $q=0.5$ a.u. and $\w_{0}=0.5$ a.u.,
is switched on in the quantum well. 
In the numerical simulations we 
set the propagation window between $x=-1.2$ and $x=1.2$ 
a.u. (which coincides with the static potential well) and choose a 
lattice spacing $\D x=0.024$ a.u.. The 
occupied part of the continuum spectrum 
is discretized with 100 $k$-points between 0 and 
$k_{\rm F}=\sqrt{2\ve_{\rm F}}$.

\begin{figure}[htbp]
\includegraphics*[width=.47\textwidth]{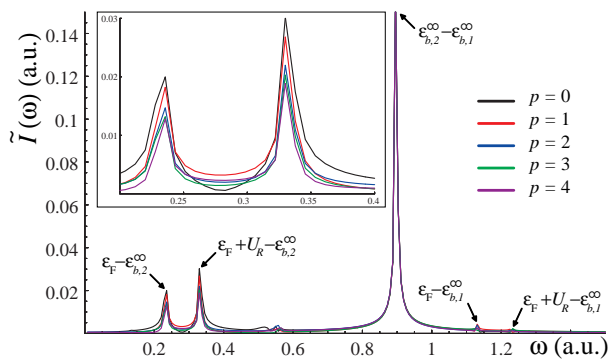}
\caption{Modulus of the discrete Fourier transform of the current for 
zero driving and $U_{R}=0.1$ a.u.. The inset shows a magnification of 
the region with bound-continuum transitions. Different curves 
correspond to different time intervals.}
\label{transientdc}
\end{figure}

Let us first consider the biased system with no driving, i.e., $U_{1}=0$. We 
propagate the (non-interacting) many-body state from $t=0$ to $t=1400$ a.u. using a time 
step $2\d=0.05$ a.u., and calculate the current $I(t)$ at the center of the quantum 
well. As in the examples of Sections \ref{archscrew} and 
\ref{nanopump}, one observes a first transient behavior which lasts for few tens 
of atomic units. However, after this first normal transient a second transient 
regime sets in. In Fig. \ref{transientdc} we plot the modulus of the discrete Fourier 
transform of the current 
\begin{equation}
    \tilde{I}(\w_{k})=2\d\sum_{n=n_{p}}^{n_{p}+N_{0}}I(2n\d)
    e^{-i\w_{k} n\d },
    \quad \w_{k}=\frac{2\p k}{N_{0}\d}
    \label{dftc}
\end{equation}
for $n_{p}=(4+2p)\cdot 10^{3}$, $p=0,1,2,3,4$, and $N_{0}=16\cdot 10^{3}$
(corresponding to the time intervals $t\in(t_{p},t_{p}+T_{0})$ with 
$t_{p}=(2+p)\times 100$ a.u. and $T_{0}=800$ a.u.). Besides the zero-frequency 
peak (not shown) due to the non vanishing dc current, the structure of $\tilde{I}(\w)$ 
has five more peaks.
Below we discuss the physical origin of these extra peaks and show 
that they are related to different kinds of internal
transitions.

We first observe that the biased system has two bound states 
with energy $\ve_{b,1}^{\inf}=-1.032$ a.u. and $\ve_{b,2}^{\inf}=-0.133$ 
a.u. (slightly different from the bound-state energies of the unbiased 
system). The first and the last two peaks occur at the same frequency
of the bound-continuum transitions 
$\ve_{b,i}^{\inf}\rightarrow \ve_{\rm F}$, and 
$\ve_{b,i}^{\inf}\rightarrow \ve_{\rm F}+U_{R}$, with $i=1,2$. These 
sharp structures (mathematically stemming from the discontinuity of the 
zero-temperature Fermi distribution function) 
give rise to long-lived oscillations of the total current and density.
Such an oscillatory transient regime dies off slowly as $1/t$. The power-low behavior can 
also be seen in the inset of Fig. 
\ref{transientdc}, where a magnification of the region with transitions from the 
weakly bound electron to the two continua is displayed. 
Denoting with $R_{p}$ the product between the height of the second peak 
and the propagation time $t_{p}+T_{0}$ we have found $R_{2}=26.305$ a.u., 
$R_{3}=26.307$ a.u., and $R_{4}=26.328$ a.u., which is in fairly good 
agreement with the expected $1/t$ behavior.
Therefore, the hight of the peaks decreases with increasing 
$t_{p}$ and approaches zero in the limit $t_{p}\ra\inf$. On the 
contrary, the sharp peak at $\w=\ve_{b,1}^{\inf}-\ve_{b,2}^{\inf}$ 
(bound-bound transition) remains unchanged with increasing $t_{p}$. 
The oscillations of the bound-bound transition do not die off, 
in agreement with the findings of Refs. \onlinecite{ds.2006,s.2006}.
We emphasize that these latter oscillations are an intrinsic property 
of the biased system and have nothing to do with external drivings.

\begin{figure}[htbp]
\includegraphics*[width=.47\textwidth]{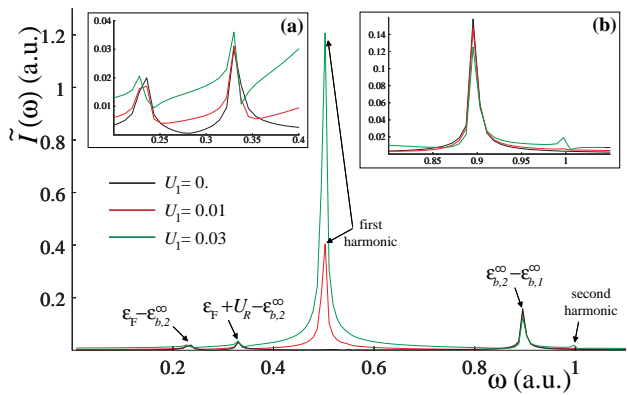}
\caption{Modulus of the discrete Fourier transform of the current for 
the biased quantum well ($U_{R}=0.1$ a.u.) perturbed by the travelling 
wave $U(x,t)=U_{1}\sin(qx-\w t)$, with $q=0.5$ a.u. and $\w=0.5$ a.u..
Three different amplitudes $U_{1}=0.00,\,0.01,\,0.03$ a.u. are 
considered. Inset (a) displays a magnification of 
the region with bound-continuum transitions. Inset (b) shows a 
magnification of the region with the bound-bound transition and the 
second-harmonic peak.}
\label{transientac}
\end{figure}

Having discussed the behavior of the system which is biased but not 
driven, we now 
study transient regimes in the biased {\em and} driven system, i.e., 
$U_{1}\neq 0$. Using the same numerical parameters as in the previous 
example we evolve the (non-interacting) many-body state from $t=0$ to $t=1200$ a.u. 
with a time step $2\d=0.05$ a.u.. In Fig. \ref{transientac} we plot 
the discrete Fourier transform of the current calculated in the middle 
of the quantum well for different amplitudes of the travelling wave  
$U_{1}=0.00,\,0.01,\,0.03$ a.u.. The time interval used to evaluate 
$\tilde{I}(\w)$ is from $t=200$ a.u. to $t=1200$ a.u.. As expected, 
$\tilde{I}(\w)$ has a well pronounced peak at the driving frequency 
(first harmonic). Increasing the amplitude of the driving field 
the height of the first-harmonic peak increases and higher-order 
harmonic peaks become visible (breakdown of linear response theory). 
This is clearly shown in inset (b) where the second-harmonic peak is 
visible for $U_{1}=0.03$ a.u. but not for $U_{1}=0.01$ a.u.. 
The structure of $\tilde{I}(\w)$ has also other peaks at frequencies which 
are not commensurable with the driving frequency. Such peaks are due to the 
presence of bound states in the biased-only system. In inset (a) we 
show a magnification of the region with bound-continuum transitions. 
The driving field broadens the peak-structure, thus 
speeding up the power-law transient regime. The shape of the 
bound-bound transition 
is displayed in inset (b). The height of the peak decreases with 
increasing amplitudes and the transition changes 
from an infinitely long-lived excitation to an excitation with a 
finite life time. Let $m_{s}$ be the smallest integer for which
$m_{s}\w_{0}>\max(|\ve_{b,1}^{\inf}|,|\ve_{b,2}^{\inf}|)$; for small amplitudes 
the life time is proportional to 
$(1/U_{1}^{2})^{m_{s}}$ according to the following reasoning. The retarded Green's 
function in region $C$ can be written in terms of the embedding 
self-energy of Eq. (\ref{rase}) and the Floquet self-energy 
$\bgS^{\rm R}_{\rm ac}$ of Eq. (\ref{acdef}). The Floquet self-energy 
generates replica of the continuous spectrum which are shifted by 
multiple integers of $\w_{0}$ and contributes to 
the imaginary part of the Green's function, $\bG^{\rm R}$. The leading-order 
contribution of the $m$-th replica to ${\rm Im}\,\bG^{\rm R}$ 
scales like $(U_{1}^{2})^{m}$. Therefore, bound-state 
simple poles of $\bG^{\rm R}$ get embedded in the continuum spectrum of some of the replica and aquire 
an imaginary part proportional to $(U_{1}^{2})^{m}$, with $m$ the 
order of the replica. The leading-order contribution to the life-time 
of the bound-bound excitation is then proportional to $(1/U_{1}^{2})^{m_{s}}$.

In conclusion, we have shown that the biased and driven quantum well has a very rich 
transient structure. This is due to the presence of bound states 
which can substantially delay  the development of the Floquet regime.

\section{Conclusions and outlooks}
\label{conc}

Time-dependent gate voltages can be used to generate a net 
current between unbiased electrodes in nanoscale junctions. 
Most works focus on periodic drivings
for which Floquet-based approaches provide a powerful 
machinery to investigate the long-time behavior 
of the system. Combining Floquet theory with nonequilibrium 
Green's functions techniques we obtained a general formula for the average 
current of monochromatically driven systems in terms of inelastic transmission 
probabilities. The case of polychromatic drivings, which has received 
scarce attention so far, is analytically more complicated and computationally 
rather costly. 

In this work we proposed an alternative approach which can deal 
with monochromatic, polychromatic and nonperiodic drivings.
The computational cost is independent of the particular 
time dependence of the driving potential. As an extra bonus we can investigate
how the transient behavior depends on the initial state and 
on the details of the switching process. The basic idea is to calculate the 
time-dependent density and current from the time-evolved (non-interacting) 
many-particle 
state. This amounts to solving a single-particle Schr\"odinger equation 
for each occupied eigenstate of the unperturbed system. We have 
given full implementation details of the time-propagation algorithm
and discussed its performance. The generalization to 
two- or three-dimensional reservoirs can be worked out following the 
general lines of Ref. \onlinecite{ksarg.2005} and its implementation 
is in progress.

We illustrated our scheme in one-dimensional structures. First we 
studied pumping through a single barrier, and showed that the electrons 
are dragged by the travelling wave and move in pockets. Second we 
studied pumping in semiconducting structures, and  
investigated the phenomenon of current inversion. In both examples 
the Floquet algorithm of Appendix \ref{accurr} is used for benchmarking 
the long-time limit of the real-time simulations and we have found an excellent 
agreement between the two approaches. Finally, we considered pumping through a 
quantum well connected to biased reservoirs. The aim of this latter 
example is to show the existence of a long-lived transient regime in 
rather common physical systems. The transient oscillations are 
explained in terms of bound-bound transitions and bound-continuum 
transitions. These oscillations usually have frequencies which are not 
commensurable with the driving frequency and are therefore not 
described by the initial Floquet assumption. 

The present work opens the path towards systematic studies of nanoscale 
devices as it is not restricted to linear response theory and can cope 
with general time-dependent as well as spatial perturbations. Our approach 
can also be extended in a natural way to describe more complicated physical 
systems. The effects of electron correlation may be included within the 
framework of time-dependent density functional theory\cite{ksarg.2005} 
by using present exchange-correlation density functionals as well as orbital 
dependent ones. Second, the scheme can be upgraded to cope with 
three-dimensional reservoirs. This is computationally more demanding but 
clearly will pay back in our understanding of non-equilibrium dynamical 
phenomena in nanoconstrictions. 

Highlighting different physical phenomena, our idea of real-time evolution of 
open quantum systems may also be used to address questions such as 
time-dependent spin transport, current fluctuations and shot noise, 
optimal control of devices for quantum information processing,
\cite{rcwrg.2006} the role of superconducting leads, heat transport, etc.. 
In particular, the design of fast, integrated, optoelectronic nanodevices  
clearly requires the proper description of dynamical effects 
(relaxation, decoherence, etc.) on a microscopic level.
Problems related to current induced heating and electromigration
should also be addressed,\cite{vsa.2006,dvp.2000,hbf.2004,mrt.2006} and 
one might need to go beyond the classical treatment of the ionic motion as 
it fails in describing Joule heating.\cite{hbftm.2004,hbfts.2004,bhst.2005}
The present work is a small step towards those ambitious goals, adding
the physics of time-dependent phenomena to the world of steady-state 
effects in quantum transport.

\begin{acknowledgments}
We thank E. Khosravi, C. Verdozzi and H. Appel for useful discussions.
This work was supported in part by the Deutsche Forschungsgemeinschaft,
DFG programme SFB658, the EU Research and Training Network 
EXCITING, the EU Network of Excellence NANOQUANTA (NMP4-CT-2004-500198), 
the SANES project (NMP4-CT-2006-017310), the DNA-NANODEVICES 
(IST-2006-029192), the 2005 Bessel research award of the Humboldt 
Foundation, and the BSC (Barcelona Mare Nostrum Center).
\end{acknowledgments}

\appendix

\section{Current formula}
\label{accurr}

The dc kernel $\bQ_{\a,\rm dc}$ in Eq. (\ref{kerneldc}) is given by 
the sum of two terms, both containing an integral over energy $\w$. 
Consequently, also the total dc current can be expressed as the sum 
of two terms. From Eq. (\ref{currtt}) it is straightforward to 
obtain
\begin{equation}
    I_{\a,\rm dc}=I_{\a}^{(1)}+I_{\a}^{(2)},
\end{equation}
with
\begin{equation}
    I_{\a}^{(1)}=-2\int\frac{\dr\w}{2\p}
    f_{\a}(\w)\,{\rm Im}\,{\rm Tr}
    \left[\bgG_{\a}(\w)\bG_{0}(\w)\right],
    \label{dc1}
\end{equation}
and
\begin{eqnarray}
    I_{\a}^{(2)}=-\int\frac{\dr\w}{2\p}
    \sum_{\b}f_{\b}(\w)\sum_{m}
    \quad\quad\quad\quad\quad
    \quad\quad\;\,
    \nonumber \\ 
    \times
    {\rm Tr}
    \left[\bG_{m}(\w)\bgG_{\b}(\w)\bG^{\dag}_{m}(\w)\bgG_{\a}(\w-m\w_{0})
    \right].
        \label{dc2}
\end{eqnarray}

Let us focus on the coeffiecients $\bG_{m}$ and derive a 
recursive scheme to calculate them. We write the Hamiltonian 
$\bH_{CC}(t)$ as the sum of a static, $\bH_{CC}^{0}$, and 
periodic, $\bU_{CC}(t)$, term and expand the latter in Fourier modes
\begin{equation}
    \bU_{CC}(t)=\sum_{n}\bU_{n}e^{in\w_{0}t},\quad\quad
    \bU_{n}=\bU^{\dag}_{-n}.
    \label{uexp}
\end{equation}
We also define the Green's function $\bg$ as the projection onto region $C$ of the Green's 
function of the system which is biased but not driven, i.e., 
$\bU_{CC}(t)=0$. The Green's function $\bg$ depends only 
on the difference between its time arguments and can be expanded as 
follows
\begin{equation}
    \bg^{\rm R}(t;t')=\sum_{m}\int\frac{\dr\w}{2\p}\bg^{\rm R}_{m}(\w)
    e^{-i\w(t-t')+im\w_{0}t'},
    \label{lgexp}
\end{equation}
where the only non-vanishing coefficient of the expansion is 
$\bg^{\rm R}_{0}(\w)$ and reads
\begin{equation}
    \bg^{\rm R}_{0}(\w)=\frac{1}
    {\w{\bf 1}_{C}-\bH_{CC}^{0}-
    \bgS^{\rm R}(\w)},
\end{equation}
with ${\bf 1}_{C}$ the unit matrix in region $C$ and $\bgS^{\rm R}$ 
the retarded embedding self-energy of Eq. (\ref{rase}).
Inserting the above expansions into the Dyson equation 
\begin{equation}
    \bG^{\rm R}(t;t')=\bg^{\rm R}(t;t')+\int \dr\bar{t}\;
    \bg^{\rm R}(t;\bar{t})\bU_{CC}(\bar{t})\bG^{\rm R}(\bar{t};t),
\end{equation}
we find a set of linear equations for the coefficients $\bG_{m}$
\begin{equation}
    \bG_{m}(\w)=\d_{m,0}\bg_{m}(\w)+\bg_{m}(\w)\sum_{n}
    \bU_{n}\bG_{m-n}(\w),
    \label{geitre}
\end{equation}
where we have used the short-hand notation $\bg_{m}(\w)=\bg^{\rm 
R}_{0}(\w-m\w_{0})$ (the $\bg_{m}$ should not to be confused with the expansion 
coefficient $\bg_{m}^{\rm R}$ of Eq. (\ref{lgexp}); the latter is zero 
for all $m\neq 0$). For arbitrary periodic drivings the 
solution of Eq. (\ref{geitre}) is computationally very hard. In the 
following we specialize to the monochromatic case and describe a 
feasible numerical scheme to calculate the $\bG_{m}$'s.

For monochromatic drivings, 
$\bU_{n}=\d_{n,1}\bU_{+}+\d_{n,-1}\bU_{-}$, the algebraic system in 
Eq. (\ref{geitre}) simplifies to (understanding the quantities as 
function of $\w$)
\begin{equation}
    \bG_{m}=\bg_{m}\left[\d_{m,0}+
    \bU_{+}\bG_{m-1}+
    \bU_{-}\bG_{m+1}\right],
    \label{tridsys}
\end{equation}
which is a tridiagonal system. In matrix form Eq. (\ref{tridsys}) 
reads
\begin{widetext}
\begin{equation}
	\left[
	\begin{array}{cccc|c|cccc}
	    &  &  &  & \vdots &  &  &  & \\ 
	    &  & \underline{\bM}^{(-)} &  & 0 &  & \underline{{\bf 0}} &  & \\ 
	     &  &  &  & 0 &  &  &  & \\ 
	      &  &  &  & -\bg_{-1}\bU_{-} &  &  & & \\ \hline      
	    \ldots & 0 & 0 & -\bg_{0}\bU_{+} & {\bf 1}_{C} & -\bg_{0}\bU_{-} & 0 & 0 & \ldots \\ \hline
	     &  &  & & -\bg_{1}\bU_{+} & &  &  &  \\ 
	     &  &  &  & 0 &  &  &  & \\ 
	     &  & \underline{{\bf 0}} &  & 0 &  & \underline{\bM}^{(+)} &  & \\ 
	     &  &  &  & \vdots &  &  &  & \\ 
	\end{array}
	\right]
	\left[
	\begin{array}{c}
	    \vdots \\
	    \bG_{-3} \\
	    \bG_{-2} \\
	    \bG_{-1} \\ \hline
	    \bG_{0} \\ \hline
	    \bG_{1} \\
	    \bG_{2} \\
	    \bG_{3} \\
	    \vdots 
	\end{array}
	\right]
	=
	\left[
		\begin{array}{c}
	    \vdots \\
	    0 \\
	    0 \\
	    0 \\ \hline
	    \bg_{0} \\ \hline
	    0 \\
	    0 \\
	    0 \\
	    \vdots 
	\end{array}
	\right]
\end{equation}
where $\underline{{\bf 0}}$ is the null matrix and 
the matrices $\bM^{(\pm)}$ read
\begin{equation}
    \underline{\bM}^{(-)}=
    \left[
    \begin{array}{ccccc}
	 & \ddots\quad\quad\quad\quad & \ddots\quad\quad\quad & 0 & 0\\
	\ldots & -\bg_{-3}\bU_{+} & {\bf 1}_{C} & -\bg_{-3}\bU_{-} & 0 \\
	\ldots & 0 & -\bg_{-2}\bU_{+} & {\bf 1}_{C} & -\bg_{-2}\bU_{-} \\
	\ldots & 0 & 0 & -\bg_{-1}\bU_{+} & {\bf 1}_{C}
    \end{array}
    \right],
\end{equation}    
\begin{equation}
    \underline{\bM}^{(+)}=
    \left[
    \begin{array}{ccccc}
	 {\bf 1}_{C} & -\bg_{1}\bU_{-} & 0  & 0 & \ldots \\
	 -\bg_{2}\bU_{+}  & {\bf 1}_{C} & -\bg_{2}\bU_{-} & 0 & \ldots \\
	 0 & -\bg_{3}\bU_{+} & {\bf 1}_{C} & -\bg_{3}\bU_{-} & \ldots \\
	 0 & 0 & \quad\quad\quad\ddots & \quad\quad\quad\quad\ddots &	
    \end{array}
    \right].
\end{equation}    
Let $\bM^{-1}_{-}$, $\bM^{-1}_{+}$ be the bottom-right block of the inverse of 
$\underline{\bM}^{(-)}$ and the top-left block of the inverse of 
$\underline{\bM}^{(+)}$ respectively. The coefficient $\bG_{\pm 1}$ 
can be expressed in terms of $\bM^{-1}_{\pm}$ according to
\begin{equation}
\bG_{\pm 1}=\bM^{-1}_{\pm}\bg_{\pm 1}\bU_{\pm}\bG_{0}.
\end{equation}
Substituting this result into Eq. (\ref{tridsys}) with $m=0$, one obtains a closed 
equation for $\bG_{0}$
\begin{equation}    
    \bG_{0}=\bg_{0}+\bg_{0}\sum_{\pm}\bU_{\mp}
    \bM^{-1}_{\pm}\bg_{\pm 1}\bU_{\pm}\bG_{0}.
    \label{g0rozza}
\end{equation}
Exploiting the tridiagonal block-structure of 
$\underline{\bM}^{(\pm)}$ we can express the matrices $\bM^{-1}_{\pm}$ 
as a continued matrix fraction
\begin{equation}
    \bM^{-1}_{\pm}=\frac{{\bf 1}_{C}}
    {{\bf 1}_{C}-\bg_{\pm 1}\bU_{\mp}\frac{\mbox{${\bf 1}_{C}$}}
    {\mbox{${\bf 1}_{C}-\bg_{\pm 2}\bU_{\mp}$}\frac{\mbox{${\bf 1}_{C}$}}
    {\quad\ddots}\mbox{$\bg_{\pm 3}\bU_{\pm}$}}\bg_{\pm 2}\bU_{\pm}}
    =\frac{{\bf 1}_{C}}
    {\bg_{\pm 1}^{-1}-\bU_{\mp}\frac{\mbox{${\bf 1}_{C}$}}
    {\mbox{$\bg_{\pm 2}^{-1}-\bU_{\mp}$}\frac{\mbox{${\bf 1}_{C}$}}
    {\quad\ddots}\mbox{$\bU_{\pm}$}}\bU_{\pm}}\bg_{\pm 1}^{-1},
\end{equation}
which is equivalent to solving the 
following recursive relations (remaking explicit the dependence on $\w$)
\begin{equation}
   \bM_{\pm}^{-1}(\w)= \bH_{\pm,1}^{-1}(\w)\bg_{\pm 1}^{-1}(\w).
\end{equation}
and
\begin{equation}
    \bH_{\pm,m}^{-1}(\w)=\frac{{\bf 1}_{C}}{\bg^{-1}_{\pm m}(\w)-
    \bU_{\mp}\,\bH^{-1}_{\pm,m+1}(\w)\,\bU_{\pm}}=
    \frac{{\bf 1}_{C}}{\left(\w\mp m\w_{0}\right){\bf 1}_{C}-\bH^{0}_{CC}-\bgS^{\rm R}(\w\mp m\w_{0})-
    \bU_{\mp}\,\bH^{-1}_{\pm,m+1}(\w)\,\bU_{\pm}}.
    \label{recre}
\end{equation}
Introducing the ac self-energy, 
\begin{equation}
    \bgS^{\rm R}_{\rm 
    ac}(\w)=\sum_{\pm}\bU_{\mp}\,\bH^{-1}_{\pm,1}(\w)\,\bU_{\pm},
    \label{acdef}
\end{equation}
which accounts for the interaction between the electrons and the ac 
driving field, we can rewrite the 
solution for $\bG_{0}$ in Eq. (\ref{g0rozza}) as
\begin{equation}
    \bG_{0}^{-1}(\w)=\bg_{0}^{-1}(\w)-\bgS^{\rm R}_{\rm ac}(\w)
    =
    \w{\bf 1}_{C}-\bH_{CC}^{0}-
    \bgS^{\rm R}(\w)-\bgS^{\rm R}_{\rm ac}(\w).
\end{equation}

In our implementation we have solved the above recursive relations by
truncating the hierarchy. For some $m=m_{\rm max}$ we set
$\bH_{\pm,m_{\rm max}}(\w)=\bg^{-1}_{\pm m_{\rm max}}(\w)$, and 
calculate all the $\bH_{\pm,m}(\w)$ with $m<m_{\rm max}$ according to 
Eq. (\ref{recre}). The convergence of the result can be tested by 
increasing $m_{\rm max}$. Typically, the smaller  $\w_{0}$ 
the larger one has to choose $m_{\rm max}$ to achieve convergence. Once the 
matrix $\bG_{0}$ has been calculated, the matrices $\bG_{m}$ with $m\neq 0$ 
are easily obtained from
\begin{equation}
    \bG_{\pm m}(\w)=\bH^{-1}_{\pm,m}(\w)\bU_{\pm}\bG_{\pm (m-1)}(\w),
    \quad m>0.
    \label{recgm}
\end{equation}

Having explicit equations for the $\bG_{m}$'s, we now show how to 
express the total dc current in terms of inelastic transmission probabilities. 
To calculate the contribution $I_{\a}^{(1)}$ in Eq. (\ref{dc1}) we need 
to evaluate the imaginary part of 
${\rm Tr}\left[\bgG_{\a}\bG_{0}\right]$. Using the identity
\begin{equation}
    \bG_{0}-\bG_{0}^{\dag}=\bG_{0}^{\dag}
    \left(\bgS^{\rm R}-[\bgS^{\rm R}]^{\dag}+\bgS^{\rm R}_{\rm ac}-
    [\bgS^{\rm R}_{\rm ac}]^{\dag}\right)\bG_{0},
\end{equation}
we find
\begin{equation}
    {\rm Im}\,{\rm Tr}\left[\bgG_{\a}\bG_{0}\right]=
    \frac{1}{2i}{\rm Tr}
    \left[\bgG_{\a}\left(\bG_{0}-\bG_{0}^{\dag}\right)\right]
    =
    -\frac{1}{2}{\rm Tr}\left[\bgG_{\a}\bG_{0}^{\dag}\left(\bgG+\bgG_{\rm 
    ac}\right)\bG_{0}\right],
\end{equation}
where we have defined $\bgG=\bgG_{L}+\bgG_{R}=i\left(\bgS^{\rm R}-[\bgS^{\rm 
R}]^{\dag}\right)$ and $\bgG_{\rm ac}=i\left(\bgS^{\rm R}_{\rm ac}-
[\bgS^{\rm R}_{\rm ac}]^{\dag}\right)$. From the recursive relation 
(\ref{recgm}) and the definition of 
$\bgS^{\rm R}_{\rm ac}$ in Eq. (\ref{acdef}) we have
\begin{equation}
    \bG_{0}^{\dag}\bgS^{\rm R}_{\rm ac}\bG_{0}=
    \sum_{\pm}\bG_{0}^{\dag}\bU_{\mp}\bH^{-1}_{\pm,1}\bU_{\pm}\bG_{0}
    =
    \sum_{\pm}\bG_{\pm 1}^{\dag}\bH^{\dag}_{\pm,1}\bG_{\pm 1},
\end{equation}
and hence
\begin{equation}
\bG_{0}^{\dag}\bgG_{\rm ac}\bG_{0}=i
\sum_{\pm}\bG_{\pm 1}^{\dag}
\left(\bH^{\dag}_{\pm,1}-\bH_{\pm,1}\right)\bG_{\pm 1}.
\label{fstfr}
\end{equation}
Next, we use the recursive relations (\ref{recre}) and find
\begin{equation}
\bH^{\dag}_{\pm,1}(\w)-\bH_{\pm,1}(\w)=
-i\bgG(\w\mp\w_{0})+
\bU_{\mp}
\left(\bH^{-1}_{\pm,2}(\w)-[\bH_{\pm,2}^{-1}(\w)]^{\dag}\right)\bU_{\pm}.
\end{equation}
Inserting this result into Eq. (\ref{fstfr}) yields
\begin{equation}
\bG_{0}^{\dag}(\w)\bgG_{\rm ac}(\w)\bG_{0}(\w)=
\sum_{\pm}\bG_{\pm1}^{\dag}(\w)
\bgG(\w\mp \w_{0})\bG_{\pm 1}(\w)
+i\sum_{\pm}\bG_{\pm 1}^{\dag}(\w)\bU_{\mp}
\left([\bH_{\pm,2}^{-1}(\w)]^{\dag}-\bH^{-1}_{\pm,2}(\w)\right)
\bU_{\pm}
\bG_{\pm 1}(\w).
\end{equation}
The second term on the r.h.s. can be expressed in terms of $\bG_{\pm 2}$ 
with the help of Eq. (\ref{recgm}). In doing so we obtain a first 
term given by $\sum_{\pm}\bG_{\pm 2}^{\dag}(\w)
\bgG(\w\mp 2\w_{0})\bG_{\pm 2}(\w)$, and a second term that can be 
expressed in terms of  $\bG_{\pm 3}$. Iterating {\em ad infinitum} we 
end up with the following expression
\begin{equation}
\bG_{0}^{\dag}(\w)\bgG_{\rm ac}(\w)\bG_{0}(\w)=
\sum_{m>0}\sum_{\pm}\bG_{\pm m}^{\dag}(\w)
\bgG(\w\mp m\w_{0})\bG_{\pm m}(\w),
\end{equation}
and therefore
\begin{equation}
{\rm Im}\,{\rm Tr}\left[\bgG_{\a}(\w)\bG_{0}(\w)\right]
=-\frac{1}{2}\sum_{m}{\rm Tr}\left[\bgG_{\a}(\w)\bG_{m}^{\dag}(\w)
\bgG(\w-m\w_{0})\bG_{m}(\w)\right].
\end{equation}
\end{widetext}
Substituting this result back into Eq. (\ref{dc1}) and performing the 
sum $I_{\a}^{(1)}+I_{\a}^{(2)}$, with $I_{\a}^{(2)}$ from Eq. 
(\ref{dc2}), we obtain the total dc current in terms of inelastic 
transmission probabilities [see Eq. (\ref{dcac})]. The above derivation is 
based on nonequilibrium Green's functions, and generalizes a previous 
derivation\cite{a.2005} to central regions of dimension larger than one.

\end{document}